\documentclass[aps,twocolumn,floatfix,bibnotes]{revtex4}

\usepackage{here}

\usepackage[dvips]{graphicx}

\newcommand\pictc[5]{\begin{figure}
                       \centerline{
                       \includegraphics[width=#1\columnwidth]{#3}}
                   \protect\caption{\protect\label{fig:#4} #5}
                    \end{figure}            }
\newcommand\pict[4][1]{\pictc{#1}{!tb}{#2}{#3}{#4}}
\newcommand\rpict[1]{\ref{fig:#1}}

\newcommand\leqt[1]{\protect\label{eq:#1}}
\newcommand\reqtn[1]{\ref{eq:#1}}
\newcommand\reqt[1]{(\reqtn{#1})}

\newcounter{Fig}

\begin{document}

\begin{sloppy}

\title{Suppression of left-handed properties in disordered metamaterials}

\author{Alexander A. Zharov$^{1,2}$}
\author{Ilya V. Shadrivov$^1$}
\author{Yuri S. Kivshar$^1$}

\affiliation{$^1$Nonlinear Physics Centre, Research School of
Physical Sciences and Engineering, Australian National University,
Canberra ACT 0200, Australia\\
$^2$ Institute for Physics of Microstructures,
Russian Academy of Sciences, Nizhny Novgorod 603950, Russia}

\begin{abstract}
We study the effect of disorder on the effective magnetic response
of composite left-handed metamaterials and their specific
properties such as negative refraction. We show that relatively
weak disorder in the split-ring resonators can reduce and even completely eliminate  the frequency domain where the composite material demonstrates the left-handed properties. We introduce the concept of the order parameter to describe novel
physics of this effect.
\end{abstract}

\maketitle

Recently fabricated composite structures~\cite{Smith:2000-4184:PRL,Bayindir:2002-120:APL,Parazzoli:2003-107401:PRL} open an unique
possibility to study experimentally the properties of the so-called left-handed
metamaterials~\cite{Veselago:1967-517:UFN}. Specifically, the composite materials created by arrays of wires
and split-ring resonators (SRRs) are known to possess a negative
real part of the magnetic permeability and negative dielectric
permittivity in the microwave range~\cite{Smith:2000-4184:PRL,Bayindir:2002-120:APL,Parazzoli:2003-107401:PRL}, and they demonstrate many unusual properties predicted theoretically long time
ago~\cite{Veselago:1967-517:UFN}, including negative refraction,
inverse light pressure, reverse Doppler and Vavilov-Cherenkov
effects, etc. Moreover, many ideas to create such left-handed metamaterials
for optical frequencies are currently under study (see,
e.g., Ref.~\cite{Podolskiy:2003-735:OE}).

The metamaterials created by arrays of metallic wires and
microwave resonators are designed to operate for the wavelengths
much larger then the period of the structure, and their
simultaneously negative dielectric permittivity and magnetic
permeability occur in some finite frequency range. Within the
effective medium approximation, dielectric
permittivity becomes negative in a relatively large frequency
domain due to a linear response of wires, whereas magnetic permeability of the structure can become negative in a relatively narrow frequency domain. It is this specific frequency domain where the induced magnetic momentum of each resonator is directed opposite to the external magnetic field being strong
enough to produce collectively negative values of the effective
magnetic permeability.

In this Letter, we study the effect of weak disorder in the
structure parameters on the magnetic response of composite and the existence of the frequency domain where the composite material exhibits the specific left-handed properties such as negative refraction. We show that even
relatively small variation of the SRR parameters can result in the
dramatic decrease of the size of the left-handed frequency domain
and, above a certain threshold value, disorder can suppress and
even eliminate completely the left-handed properties of the
metamaterial.

We study a composite structure made of metallic wires and
single-ring microwave resonators, as shown schematically in
Fig.~\rpict{str_psi}(a). This model is qualitatively similar to
the model of double-ring resonators usually studied in the theory
of left-handed composite media (see, e.g.,
Ref.~\cite{Markos:2002-36622:PRE} and reference therein). Within
the effective medium approximation, the dielectric permittivity of
this structure can be
calculated~\cite{Pendry:1999-2075:ITMT,Zharov:2003-37401:PRL} and
presented in the form
\begin{equation}
\leqt{eps}
\epsilon_{\rm eff}(\omega) \approx 1 -
\frac{\omega_p^2}{\omega\left(\omega-
    i\gamma_e\right)},
\end{equation}
where $\omega_p = 2\pi c/d \ln \left(d/r_w\right)$, $d$ is the
size of the unit cell, $r_w$ is the radius of a metallic grid,
$\gamma_e = c^2/2\sigma S \ln\left(d/r_w\right)$, $\sigma$ is
conductivity of a wire, $S$ is the effective wire cross-section,
$\omega$ is the angular frequency, and $c$ is the free-space speed
of light. Usually, the effective plasma frequency,
$f_p=\omega_p/2\pi$, is between $10$ GHz and $15$ GHz.

The most intriguing properties of the left-handed composite
metamaterials are due to a SRR response which
determines effective magnetic permeability. Magnetization of
the metamaterial with three SRRs per unit cell, each of them
having the axis in the perpendicular direction with respect to others
as shown in Fig.~\rpict{str_psi}(a), can be presented in the form
(see, e.g.,
Refs.~\cite{Shadrivov:2004-46615:PRE,Schwinger:1998:ClassicalElectrodynamics}),
${\bf M} = \chi(\omega){\bf H^{\prime}}$, where
\begin{equation}
\leqt{chi}
\chi(\omega) = \frac{\eta
\omega^2}{\omega_0^2-\omega^2+i\gamma_m\omega},
\end{equation}
\begin{equation}
\leqt{eta}
\eta = \frac{\pi}{8} \left(\frac{a}{d}\right)^3 \left[\ln
(8a/r)-7/4 \right]^{-1/2}.
\end{equation}
Here, ${\bf H^{\prime}}$ is acting (microscopic) magnetic field,
$\omega_0 = \left[d_g c^2/\pi\epsilon r^2 a \left(\ln(8a/r) -7/4
\right)\right]^{1/2}$ is the SRR eigenfrequency, $a$ is a SRR radius, $\gamma_m = c^2 / 2 \sigma_r S_r \left[ \ln(8a/r) -7/4
\right]$ is damping coefficient, $d_g$ is the size of the SRR slot,
$\epsilon$ is permittivity of the dielectric infilling the SRR
slot in the structure.

Results~(\ref{eq:chi}) and (\ref{eq:eta}) are obtained under the
major assumption that all resonators in the structure are
identical. Now we consider the case when the size of the slot
$d_g$ is random and it is characterized by a statistical
distribution function. Our study can be motivated by a number of
the recent efforts to create the metamaterials with the
left-handed properties for shorter wavelengths and, for the
smaller SRR size in the structure, the key parameters would be difficult to
control in fabrication. In our structure, the parameter $\eta$
does not depend on $d_g$ [see Eq.~\reqt{eta}], and only the SRR
eigenfrequency $\omega_0$ is affected by fluctuations of $d_g$.
Then, the equation for the magnetic susceptibility \reqt{chi} can
be generalized to describe the case of randomly varying eigenfrequency,
\begin{equation}
\leqt{chi2}
\chi(\omega) = \eta \omega^2 \int_{0}^{\infty}\frac{F(X) \, dX}{X^2-\omega^2+i\gamma_m\omega},
\end{equation}
where $F(X)$ is the normalized distribution function of the SRR
eigenfrequencies, i.e. $\int_0^{\infty}F(X)\,dX = 1$. In the
standard case when all SRRs are identical, the distribution
function can be represented as $F(X) = \delta(X-\omega_0)$.

To describe the coupling between the acting magnetic field ${\bf
H^{\prime}}$ and the macroscopic magnetic field ${\bf H}$, we use
the Lorentz-Lorenz formula~\cite{Born:2002:PrinciplesOptics},
${\bf H^{\prime}} = {\bf H} +(4\pi/3) {\bf M}$, and present the
effective magnetic permeability in terms of local magnetic
susceptibility \reqt{chi2} as follows
\begin{equation}
\leqt{mu_eff} \mu_{\rm eff}(\omega) = \frac{1 +
(8\pi/3)\chi(\omega)}{1 - (4\pi/3) \chi(\omega)}.
\end{equation}
For definiteness, we consider the Lorenz-type distribution of the
SRR eigenfrequencies in the form
\begin{equation}
\leqt{distrib}
F(X) = \frac{(\Gamma/\pi)}{(X-\omega_0)^2+\Gamma^2},
\end{equation}
with a narrow width for $\Gamma \ll \omega_0$ such that the
eigenfrequencies of all resonators are close to some mean value
$\omega_0$. As a result, non-vanishing contributions to the
integral are given by the values of $X$ in the vicinity of $X =
\omega_0$, and we introduce a new variable $\Delta$, $X = \omega_0
+ \Delta$, where $|\Delta| \ll \omega_0$. We are interested in the
behavior of the magnetic susceptibility in the vicinity of
$\omega=\omega_0$, and we introduce $\omega= \omega_0 + \Omega$,
where $|\Omega| \ll \omega_0$. In this approximation,
Eq.~\reqt{chi2} can be rewritten as
\begin{equation}
\leqt{chi3}
\chi(\omega) \approx \frac{\eta}{2\pi} \omega_0 \Gamma
\int_{-\infty}^{\infty}\frac{
d\Delta}{(\Delta^2+\Gamma^2)(\Delta-\Omega+i\gamma_m/2)}.
\end{equation}
Using the contour integration, the integral in Eq.~\reqt{chi3} can
be calculated explicitly, and the expression for the magnetic
susceptibility can be obtained in the following form:
\begin{equation}
\leqt{chi4}
\chi(\Omega) = \frac{\omega_0 \eta}{2} \frac{1}{(-\Omega + i
\bar{\Gamma})},
\end{equation}
where $\bar{\Gamma} = \Gamma + \gamma_m/2$. The result
(\ref{eq:chi4}) shows that the random variation of the SRR
eigenfrequencies {\em is equivalent to some
additional losses} in the structure, and even for an ideal case of
lossless resonators, i.e., when $\gamma_m =0$, the composite
structure with random variation of the SRR frequencies possesses
effective losses. From the physical point of view, such effective
losses resemble the collisionless Landau damping in
plasmas~\cite{Lifshitz:1981:Kinetics} caused by the presence of
resonant particles and, simultaneously, it follows from the
Kramers-Kronig relations.

\pict{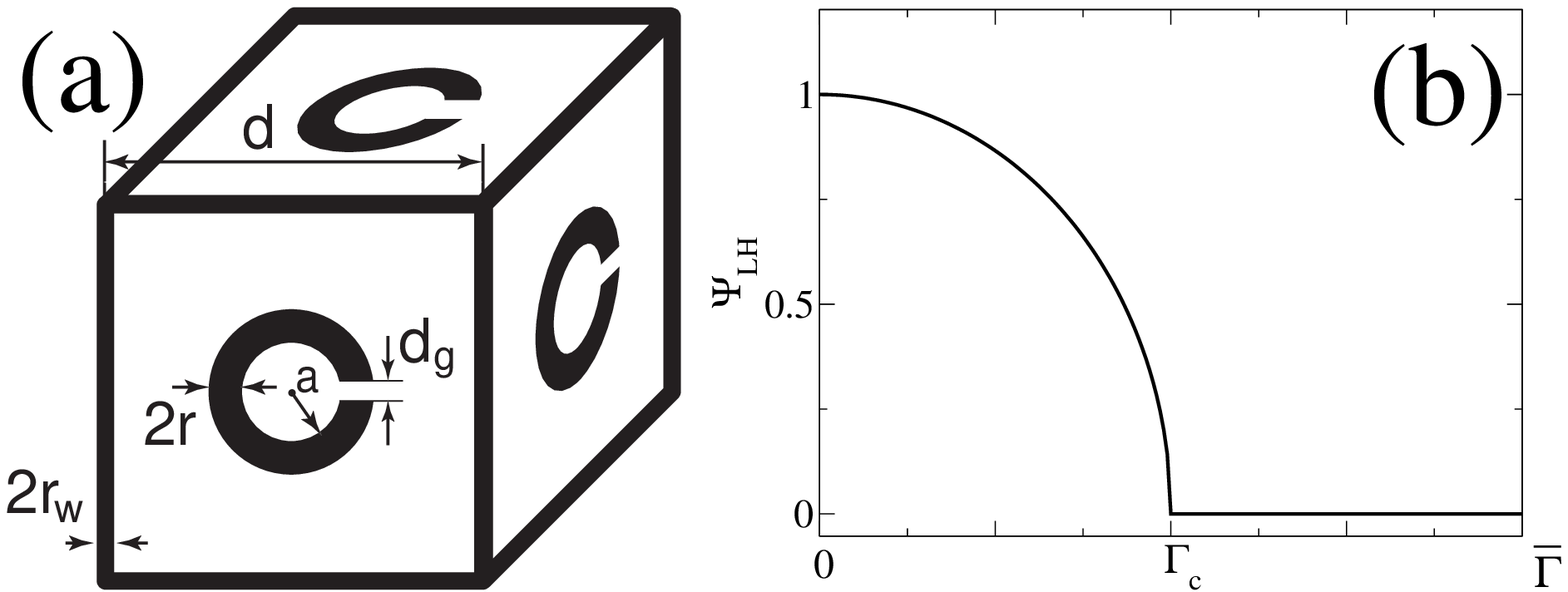}{str_psi}{(a) Schematic of the metamaterial
structure. (b) Dependence of the real part of the order parameter
$\Psi_{\rm LH}$ on the effective combined losses $\bar{\Gamma}$ in
the composite structure.}

From Eqs.~(\ref{eq:mu_eff}) and (\ref{eq:chi4}), we find the
frequency domain where the real part of magnetic permeability
becomes negative, $\Omega_1 < \Omega < \Omega_2$, where
\begin{equation}\leqt{bounds}
\Omega_{1,2} = \frac{1}{3} \Gamma_c \mp \sqrt{\Gamma_c^2 -
\bar{\Gamma}^2},
\end{equation}
and $\Gamma_c = \pi \eta \omega_0$. Thus, the width of the
frequency domain of the negative effective magnetic permeability is
\begin{equation}\leqt{LH_range}
\Delta\Omega_{\rm LH} = 2\sqrt{\Gamma_c^2 - \bar{\Gamma}^2}.
\end{equation}
The value $\Gamma_c$ has the meaning of the critical losses in the
composite structure above which the domain with negative values
of the magnetic permeability disappears, and the relative critical
parameter $\Gamma_c/\omega_0$ depends only on the structure of the
composite. Taking the characteristic values $d=0.5$cm, $a =
0.2$cm, and $r = 0.05$cm, we obtain $\Gamma_c/\omega_0 = 0.046$.
This result indicates that even for the case of lossless
resonators, i.e. $\gamma_m=0$, relative fluctuations of the SRR
eigenfrequencies cannot exceed $4.6$\%. Accordingly, if the slot
size is $d_g = 0.01$cm, (then $\omega_0/2\pi = 9.6$GHz) the
critical variation can be estimated as $\Delta d_g \sim 5\mu$m.
Existence of such a critical distribution width places strict
requirements on the manufacturing of such materials. Although for
microwaves such requirements can be easily met, they can result in
substantial experimental difficulties for shorter wavelengths. In
particular, the recent proposal to fabricate left-handed
metamaterials in optics using nanowires and $\pi$-shaped particles
~\cite{Podolskiy:2003-735:OE}, would require an accuracy better
than 5 nm, that can be challenging for the existing technology.

\pict{fig02.eps}{mu_eff}{Real part of the effective magnetic
permeability as a function of the wave frequency for
$\bar{\Gamma}/\Gamma_c = 0.2$ (solid), $\bar{\Gamma}/\Gamma_c =
0.5$ (dashed), and $\bar{\Gamma}/\Gamma_c = 1$ (dotted).}

We note that premeditative introduction of disorder does not allow
to increase the size of the left-handed frequency domain for {\em any type}
of the distribution function. Such a conclusion has a simple
physical explanation. Indeed, disorder in the eigenfrequencies for
any given SRR density results in a decrease of the effective
number of resonators, which contribute into the negative
magnetization. In the recent experiment~\cite{Chen:2004-5338:JAP},
combining s-shaped resonance particles, the authors fabricated the
metamaterial with two domains of the negative refraction. Using
the results obtained above, we may come to the conclusion that the
total size of the frequency domain with the negative magnetic
permeability in the structure fabricated in
Ref.~\cite{Chen:2004-5338:JAP} is less than it would be for the
metamaterial where all resonators are identical. Moreover, using
the resonators with several different values of the eigenfrequency can
eliminate completely the left-handed properties.

Reduction and complete suppression of the frequency domain
with negative magnetic permeability with a growth of the value of
$\bar{\Gamma}$ can also be explained in a different way. Indeed,
the left- and right-handed properties of the metamaterial can be
treated as two different ``phase states'' of the structure. Thus,
a transition from one state to the other can be interpreted as
the phase transition of the second kind. The parameter
$\bar{\Gamma}$ describes the effect of disorder in the structure, and it can be
treated as "effective temperature". Such comparison is adequate
since the effective temperature is determined as the width of
statistical frequency fluctuations including both homogeneous and
inhomogeneous line broadening. To describe different phase states
of the structure, we introduce the effective order parameter,
as is usually done in the theory of phase transitions. In the
absence of homogeneous ($\gamma_m=0$) and inhomogeneous
($\Gamma=0$) SRR line broadening (i.e. in the absence of
disorder), the metamaterial has the maximum width of the
left-handed domain, which decreases and then disappears at the
critical "effective temperature" $\Gamma_c$ termed as the
temperature of the phase transition. We use the ratio of the
left-handed frequency range to the maximum left-handed frequency
domain as the order parameter,
\begin{equation}\leqt{order_param}
\Psi_{\rm LH} = \Delta \Omega_{\rm LH}/\Delta \Omega_{\rm LH}^{\rm
(max)}= Re[1-(\bar{\Gamma}/\Gamma_c)^2]^{1/2}.
\end{equation}
so that the metamaterial possesses the left-handed properties when
$\Psi_{\rm LH} \ne 0$, i.e., below the effective
critical temperature $\Gamma_c$. Dependence of the order parameter
$\Psi_{\rm LH}$ on $\bar{\Gamma}$ is shown in
Fig.~\rpict{str_psi}(b). The real part of the magnetic
permeability as a function of the wave frequency is shown in
Fig.~\rpict{mu_eff} for different values of the ratio
$\bar{\Gamma}/\Gamma_c$. For large values of $\bar{\Gamma}/\Gamma_c$,  the frequency domain where the material possesses negative magnetic permeability is eliminated due to disorder.

The result~\reqt{order_param} suggests two ways for increasing the
width of the effective frequency domain where the composite
material possesses the left-handed properties, namely, (i)
decreasing the effective temperature $\bar{\Gamma}$, e.g. by
improving the manufacturing technology for shorter wavelengths,
and (ii) increasing the effective critical temperature $\Gamma_c$,
e.g. by a design of the resonators.

Though the concept of the order parameter $\Psi_{\rm LH}$ does
not seem to be important in the context of the problem
under study, it is expected to be useful in other (e.g.
nonlinear) problems, when it becomes dependent on the field intensity, time,
and coordinate. In this case, the order parameter will determine
the material properties in different domains, describing the
dynamics of the phase transitions between the states with left-
and right-handed properties~\cite{Zharov:2003-37401:PRL}. The
study of the nonlinear regime will be presented elsewhere.

In conclusion, we have analyzed the effect of disorder in the
composite structures which exhibit left-handed properties. In
particular, we have studied how random variation of the SRR
eigenfrequencies in the structure can change the existence of the
frequency domain where the effective dielectric permittivity and
magnetic permeability are both negative. We have demonstrated that
even relatively weak disorder in the SRR parameters can result in
a dramatic reduction of the size of the left-handed frequency domain.
More importantly, above a certain threshold value of this
disorder, the left-handed frequency domain can disappear
completely. We believe our result provide a useful guide for a
design of novel types of metamaterials operating for shorter
wavelengths where the effect of disorder is expected to be
crucially important.

A.A.Z. acknowledges a partial support from the Russian Foundation
for Basic Research (grant 05-02-16357) and thanks Nonlinear
Physics Centre at the Australian National University for
hospitality and fellowship.

\end{sloppy}
\end{document}